\documentclass[aps, twocolumn, secnumarabic, superscriptaddress, showpacs, preprintnumbers, nofootinbib]{revtex4-1}
\pdfoutput=1
\usepackage{xspace}
\usepackage{pifont}
\usepackage{soul}
\usepackage{todonotes}
\usepackage{multirow,array}
\usepackage[english]{babel}
\usepackage{amsmath,amssymb,amsfonts,bm,bbm,slashed}
\usepackage{graphicx}
\usepackage[sort&compress]{natbib}
\usepackage{xcolor}
\usepackage[normalem]{ulem}
\usepackage{hyperref}
\usepackage{cleveref}
\definecolor{red}{rgb}{1.0, 0, 0}
\usepackage{enumerate}
\usepackage{epsfig, subfigure}
\usepackage{epstopdf}
\usepackage{setspace}
\usepackage{booktabs, tabularx}
\usepackage{units}
\usepackage{tikz}
\usepackage{ctable}
\usepackage{comment}
\excludecomment{toexclude}
\includecomment{toinclude}

\usetikzlibrary{arrows,shapes}
\usetikzlibrary{trees}
\usetikzlibrary{matrix,arrows}                          % For commutative diagram
\usetikzlibrary{positioning}                            % For "abov e of=" commands
\usetikzlibrary{calc,through}                           % For coordinates
\usetikzlibrary{decorations.pathreplacing}  % For curly braces
\def\fun#1#2{\lower3.6pt\vbox{\baselineskip0pt\lineskip.9pt
  \ialign{$\mathsurround=0pt#1\hfil##\hfil$\crcr#2\crcr\sim\crcr}}}

\input epsf

\newenvironment{Eqnarray}%
         {\arraycolsep 0.14em\begin{eqnarray}}{\end{eqnarray}}
\newcommand{\be}{\begin{equation}}
\newcommand{\ee}{\end{equation}}
\newcommand{\bea}{\begin{Eqnarray}}
\newcommand{\eea}{\end{Eqnarray}}

\def\lsim{\mathrel{\raise.3ex\hbox{$<$\kern-.75em\lower1ex\hbox{$\sim$}}}}
\def\gsim{\mathrel{\raise.3ex\hbox{$>$\kern-.75em\lower1ex\hbox{$\sim$}}}}
\def\lsub#1{_{\lower 1.5pt\hbox{$\scriptstyle#1$}}}

\def\be{\begin{equation}}
\def\ee{\end{equation}}
\def\bea{\begin{eqnarray}}
\def\eea{\end{eqnarray}}

\begin{document}

\title{Higgs alignment and novel $CP$-Violating observables in two-Higgs-doublet models}

\author{
\mbox{Ian Low$^{\,a,b}$, Nausheen R. Shah$^{\,c}$, Xiao-Ping Wang$^{\,d,e}$,}
 }
\affiliation{
$^a$\mbox{{High Energy Physics Division, Argonne National Laboratory, Argonne, Illinois 60439, USA}}\\
$^b$\mbox{{Department of Physics and Astronomy, Northwestern University, Evanston, Illinois 60208, USA}} \\
$^c$\mbox{{Department of Physics and Astronomy, Wayne State University, Detroit, Michigan 48201, USA}}\\
$^d$\mbox{{School of Physics, Beihang University, Beijing 100191,China}}\\
$^e$\mbox{{Beijing Key Laboratory of Advanced Nuclear Materials and Physics, Beihang University, Beijing 100191,China}}
}

\begin{abstract}

Null results from searches for new physics at the Large Hadron Collider (LHC) tend to enforce the belief that new particles must be
much heavier than the weak scale. We undertake a systematic study of the interplay between Higgs alignment and $CP$-violation in complex two-Higgs-doublet models, which enables us to construct
a $CP$-violating scenario where new Higgs bosons are close to the weak scale after including stringent constraints from the electric dipole moment and measurements at the LHC.  In addition, we propose 
a novel potential signal of $CP$-violation in the Higgs-to-Higgs decays, $h_3\to h_2 h_1$, where $h_3, h_2$, and $h_1$ are the heaviest, second heaviest and the Standard Model-like neutral Higgs bosons, respectively.  The decay could manifest itself in triple boson final states in  $h_1h_1h_1$ and $h_1h_1Z$, which are quite
distinct and provide  unique venues for new measurements at the LHC.  

\end{abstract}

\maketitle

\section{Introduction} 

$CP$-violation (CPV) is a critical ingredient for the matter-antimatter asymmetry in the Universe  \cite{Sakharov:1967dj} and its presence is of existential significance. However,  the amount of CPV in the Standard Model (SM), via the Cabbibo-Kobayashi-Maskawa mechanism, is insufficient to generate the observed baryon asymmetry \cite{Christenson:1964fg,Aaij:2019kcg}; new sources of CPV must be present outside of the SM. A  two-Higgs-doublet model~(2HDM) \cite{Branco:2011iw} is not only one of the simplest extensions of the SM which may provide new sources for CPV \cite{Lee:1973iz,Gunion:2005ja,Haber:2012np}, but also the prototype employed in numerous more elaborate new physics models \cite{Mohapatra:1974hk}.

There is vast literature on CPV and 2HDMs. However, the majority of these studies 
focus on detecting a $CP$-even and $CP$-odd mixture in a mass eigenstate through  angular correlations or asymmetries in kinematic distributions \cite{Shu:2013uua,Chen:2014ona,Chen:2015gaa,Fontes:2015mea, Grzadkowski:2016lpv,Fontes:2017zfn,Cheung:2020ugr,Kanemura:2020ibp,Bian:2020vzc}, which requires significant experimental resources and statistics.\footnote{An exception is Ref.~\cite{,Fontes:2015xva} which proposed a combination of three different decay channels.} On the other hand, there are two major results derived from data collected at the LHC: 1) null searches for new particles beyond the SM, and 2) a SM-like 125 GeV Higgs. The first result suggests that
new particles, if present, should be much heavier than the weak scale, while the latter  
implies a dominantly $CP$-even 125 GeV Higgs.

In light of these considerations, it becomes clear that we must reevaluate the possibility of CPV  in 2HDM under the assumption of a SM-like 125 GeV Higgs, which is dubbed the alignment limit \cite{Gunion:2002zf,Delgado:2013zfa,Craig:2013hca}. Of particular interest  is the ``alignment without decoupling" limit, where new Higgs bosons could still be present near the weak scale \cite{Carena:2013ooa,Carena:2014nza,Carena:2015moc}. This has been done only under limited purview in the past \cite{Grzadkowski:2014ada, Grzadkowski:2018ohf,Kanemura:2020ibp} but we aim to achieve a comprehensive and analytical understanding. 

Specifically we emphasize there are two distinct sources of CPV  in 2HDM; in the mixing and in the decay of the Higgs bosons.
 Kinematic distributions are only sensitive to CPV in the mixing. This realization allows us to construct 
 a benchmark scenario where new Higgs bosons are not far above the weak scale, at around 500 GeV or lighter, and propose a novel signature of CPV, without recourse to angular correlations or electric dipole moment~(EDM) signals, in the Higgs-to-Higgs decay, ($h_3\to h_2h_1\to 3 h_1$), whose existence is sufficient to establish CPV  in complex
two-Higgs-doublet models (C2HDMs).\footnote{In models with additional $CP$-even scalars beyond the 2HDM, such decays may be present without CPV~\cite{Baum:2018zhf, Baum:2019pqc}. However, the mass spectrum in this case is different from that of 2HDM.}
The presence of such an observable is nontrivial, as this decay channel vanishes in the exact alignment limit. Our benchmark survives  constraints from EDMs
\cite{Baker:2006ts,Griffith:2009zz,Wang:2014hba,Andreev:2018ayy,Altmannshofer:2020shb} and  collider measurements, and could be discovered at the LHC in the near future.

%%%%%%%%%%%%%%%%%%%%%%%%%%%%%%%%%%%%%%%

\section{The Higgs Basis} 

The most general potential for a 2HDM \cite{Georgi:1978xz,Carena:2002es,Davidson:2005cw} in terms of the two hypercharge-1, $SU(2)$ doublet fields $\Phi_a=(\Phi_a^+, \Phi_a^0)^T,  a =\{1, 2\}$, is given by
\begin{align}
{\cal V}&=m^2_1 \Phi_1^\dag \Phi_1 + m^2_2 \Phi_2^\dag \Phi_2  -\left( m_{12}^2 \Phi_1^\dag \Phi_2 + {\rm H.c.}\right) \nonumber \\
& + \frac{\lambda_1}{2} (\Phi_1^\dag \Phi_1)^2+ \frac{\lambda_2}{2}(\Phi_2^\dag \Phi_2)^2 +\lambda_3 (\Phi_1^\dag \Phi_1)(\Phi_2^\dag \Phi_2)   \nonumber \\
&+\lambda_4 (\Phi_1^\dag \Phi_2)(\Phi_2^\dag \Phi_1) + \left [\frac{\lambda_5}{2} (\Phi_1^\dag \Phi_2)^2+  \lambda_6 (\Phi_1^\dag \Phi_1)( \Phi_1^\dag \Phi_2) \right.\nonumber\\
& \left.+ \lambda_7 (\Phi_2^\dag \Phi_2)( \Phi_1^\dag \Phi_2)  + {\rm H.c.} \right ]\ .\label{eq:V1}
\end{align}
We assume a vacuum preserving the $U(1)_{em}$ gauge symmetry and adopt a convention where both scalar vacuum expectation values (VEVs) are real and non-negative,
\be
\label{eq:scalarvev}
\langle \Phi_1 \rangle = \frac1{\sqrt{2}}\left( \begin{array}{c} 
                                                                                  0 \\
                                                                                  v_1
                                                                                  \end{array}\right) \ , \qquad
\langle \Phi_2 \rangle = \frac1{\sqrt{2}}\left( \begin{array}{c} 
                                                                                  0 \\
                                                                                  v_2
                                                                                  \end{array}\right) \ ,
\ee
 where  $ \sqrt{v_1^2+v_2^2}\equiv v= 246$ GeV.  It is customary to define an angle $\beta$ through $\tan\beta = {v_2}/{v_1}$.
                                                                                                                                                                
We choose to study the alignment limit  \cite{Carena:2013ooa,Carena:2014nza,Carena:2015moc} in the Higgs basis \cite{Botella:1994cs}, which is defined by two doublet fields, $H_i, i=\{1,2\}$, having the following property
\be
\label{eq:higgsdef}
\langle H_1^0 \rangle = {v}/{\sqrt{2}} \ , \qquad
\langle H_2^0 \rangle = 0 \ .
\ee
We will parametrize the Higgs basis doublets as $H_1=( G^+ ,  ( v+\phi_1^0+iG^0)/\sqrt{2})^T$ and ${H}_2=( H^+ , ( \phi_2^0+ia^0)/\sqrt{2} )^T$, where $G^+$ and $G_0$ are the Goldstone bosons. The neutral fields are $\phi_1^0$, $\phi_2^0$ and $a^0$, and the charged field is $H^+$. Moreover, our phase convention is such that $\phi_2^0$ and $a^0$ are the $CP$-even and $CP$-odd eigenstates, defined with respect to the fermion Yukawa couplings.
There is a residual $U(1)$ redundancy in the Higgs basis, labeled by $H_2\to e^{i\eta}H_2$, which leaves Eq.~(\ref{eq:higgsdef}) invariant and motivates writing the scalar potential in terms of ${\cal H}_2\equiv e^{i\eta} H_2$~\cite{Boto:2020wyf},
\begin{align}
\label{eq:finalpotential}
{\cal V}&=Y_{1} {H}_1^\dag {H}_1 +  Y_{2} {\cal H}_2^\dag {\cal H}_2  +\left( Y_{3} e^{-i\eta}  {H}_1^\dag {\cal H}_2 + {\rm H.c.}\right) \nonumber \\
& + \frac{Z_1}{2} ({H}_1^\dag {H}_1)^2+ \frac{Z_2}{2}({\cal H}_2^\dag {\cal H}_2)^2   \nonumber \\
&+Z_3 ({H}_1^\dag {H}_1)({\cal H}_2^\dag {\cal H}_2) + Z_4 ({H}_1^\dag {\cal H}_2)({\cal H}_2^\dag {H}_1) \nonumber\\
& +\left[ \frac{{Z}_5}{2} e^{-2i\eta} ({H}_1^\dag {\cal H}_2)^2+ {Z}_6 e^{-i\eta} ({H}_1^\dag {H}_1)  ( {H}_1^\dag {\cal H}_2) \right.\nonumber\\
&\left.+ {Z}_7 e^{-i\eta} ({\cal H}_2^\dag {\cal H}_2) ( {H}_1^\dag {\cal H}_2) + {\rm H.c.} \right ] \ .
\end{align}   
In the above, different choices of parameters truly represent physically distinct theories \cite{Boto:2020wyf}. The potentially complex parameters are $\{Y_3, Z_5, Z_6, Z_7\}$.

The minimization of the scalar potential gives $Y_1= - Z_1/2 v^2$ and ${Y}_3 = -  {Z}_6 v^2/2$. The first relation can be viewed as the definition of  $v$ in the Higgs basis, while the second relation implies there are only three independent complex parameters, usually taken to be $\{Z_5, Z_6, Z_7\}$.  If one can find a choice of $\eta$ such that all parameters in Eq.~(\ref{eq:finalpotential}) are real after imposing the minimization condition, the vacuum and the bosonic sector of the 2HDM is $CP$-invariant.  This can happen if and only if \cite{Lavoura:1994fv}
\be
\label{eq:cpconserve}
{\rm Im}(Z_5^* Z_6^2) = {\rm Im}(Z_5^*Z_7^2)={\rm Im}(Z_6^*Z_7) = 0 \ .
\ee
Otherwise, $CP$ invariance is broken. 

In a 2HDM the most general Higgs-fermion interactions result in tree-level flavor-changing neutral currents, which can be removed by imposing a discrete $\mathbb{Z}_2$ symmetry \cite{Glashow:1976nt,Paschos:1976ay,Georgi:1978ri}, $\Phi_1 \to \Phi_1$ and $\Phi_2 \to - \Phi_2$. In addition, the $\mathbb{Z}_2$ symmetry can be broken softly by mass terms, leading to  $\lambda_6=\lambda_7 = 0$ in Eq.~(\ref{eq:V1}). 
 
In the Higgs basis, the existence of a softly broken $\mathbb{Z}_2$ symmetry is guaranteed through the condition \cite{Haber:2015pua,Boto:2020wyf}, 
\begin{align}
\label{eq:condz2}
&(Z_1-Z_2)\left[(Z_{3}+Z_4)(Z_{6}+Z_7)^* -Z_2 Z_6^*-Z_1 Z_7^*\right.\nonumber\\
&  \left.+Z_5^*(Z_{6}+Z_7)\right]-2 (Z_{6}+Z_7)^*(|Z_6|^2-|Z_7|^2)=0\ .
\end{align}
 Equation.~(\ref{eq:condz2}) assumes $Z_{6}+Z_7\neq 0$ and $Z_1\neq Z_2$,  and eliminates two real degrees of freedom. In the end there are a total of nine real parameters in a complex 2HDM.

%%%%%%%%%%%%%%%%%%%%%%%%%%%%%%%%%%%%%%%

\section{The Alignment Limit} 

The alignment limit \cite{Gunion:2002zf,Delgado:2013zfa,Craig:2013hca} is defined by the limit where the scalar carrying the full VEV in the Higgs basis is aligned with the 125 GeV mass eigenstate \cite{Carena:2013ooa, Carena:2014nza,Carena:2015moc}, in which case the observed Higgs boson couples to the electroweak gauge bosons with  SM strength.  
The mass-squared matrix $\mathcal{M}^2$ in the $\phi^0_1-\bar{\phi}^0_2-\bar{a}^0$ 
basis, where ${\cal H}_2=( \bar{H}^+ , ( \bar{\phi}_2^0+i\bar{a}^0)/\sqrt{2} )^T$,  can be diagonalized by an  orthogonal matrix $R$ relating $\vec{\phi}=(\phi_1^0,\bar{\phi}_2^0,\bar{a}^0)^T$ to the mass eigenstates $\vec{h}=(h_3, h_2, h_1)^T$, $\vec{h}=R \cdot \vec{\phi}$ \cite{Boto:2020wyf},
\bea
&&\!\!\!\!\!\!\!\!R=R_{12}R_{13}\overline{{R}}_{23}\nonumber\\
 &&\!\!\!\!\!\!\!\!=\left(
\begin{array}{ccc}
c_{12} & -s_{12} &0\\
s_{12}  & c_{12}&0\\
0 & 0 & 1
\end{array}
\right)\!\!\!\left(
\begin{array}{ccc}
c_{13}  & 0& -s_{13}\\
0 & 1&0\\
s_{13}  &0  & c_{13}\end{array}
\right)\!\!\!\left(
\begin{array}{ccc}
1 & 0 &0\\
0 &  \bar{c}_{23} &- \bar{s}_{23}\\
0 & \bar{s}_{23}  &  \bar{c}_{23}
\end{array}
\right).
\eea
Here we have used the notation $c_{ij}=\cos \theta_{ij}$, $s_{ij}=\sin\theta_{ij}$, $\bar{c}_{23}=\cos\bar{\theta}_{23}$ and $\bar{s}_{23}=\sin\bar{\theta}_{23}$  . An important observation  is that  $\bar{\theta}_{23}$ \cite{Haber:2006ue} rotates between $\bar{\phi}_2^0$ and $\bar{a}^0$, which corresponds to the phase rotation ${\cal H}_2\to e^{i\bar{\theta}_{23}}{\cal H}_2$. Therefore the effect of the $\bar{\theta}_{23}$ rotation is to shift the $\eta$ parameter labelling the Higgs basis. In the end the combination that appears in the physical couplings is $\theta_{23}\equiv \eta+\bar{\theta}_{23}$. This motivates defining  \cite{Boto:2020wyf}
\begin{align} \label{eq:m2def}
&\widetilde{\mathcal{M}}^2\equiv\overline{R}_{23}\, \mathcal{M}^2\, \overline R_{23}^T\nonumber \\
&=v^2\left(
\begin{array}{ccc}
Z_1 & {\rm Re}[\tilde{Z}_6] & - {\rm Im}[\tilde{Z}_6]\\
 {\rm Re}[\tilde{Z}_6] &{\rm Re}[\tilde{Z}_5]+ A/v^2  & -\frac{1}{2}{\rm Im}[\tilde{Z}_5] \\
 - {\rm Im}[\tilde{Z}_6] & -\frac{1}{2}{\rm Im}[\tilde{Z}_5] &  A/v^2 
\end{array} \right) \ ,
\end{align}
where $\tilde{Z}_5=Z_5 e^{-2i\theta_{23}}$, $\tilde{Z}_{6/7}=Z_{6/7} e^{-i\theta_{23}}$, and $A=Y_2+v^2( Z_3+Z_4-{\rm Re}[\tilde{Z}_5])/2$. Alignment is achieved by the conditions ${\rm Re}[\tilde{Z_6}]={\rm Im}[\tilde{Z_6}] = 0$. 

$\widetilde{\mathcal{M}}^2$ can  be diagonalized by just two angles: $\widetilde R\, \widetilde{\mathcal{M}}^2\, \widetilde{R}^T={\rm diag}\, ( m_{h_3}^2,m_{h_2}^2,m_{h_1}^2 )$ and 
\begin{align}
\widetilde R=R_{12}R_{13}&=\left(
\begin{array}{ccc}
c_{12}c_{13} & -s_{12} & -c_{12}s_{13}\\
s_{12} c_{13} & c_{12} &-s_{12}s_{13} \\
s_{13} & 0 & c_{13}
\end{array}\right) \ ,
\label{eq:RatationMatix}
\end{align}
which relates the mass eigenbasis $(h_1, h_2, h_3)$ to the $CP$-eigenbasis $(\phi_1^0, \phi_2^0, a^0)$
\be
\left(\begin{array}{c}
\label{eq:wideRmixing1}
h_3\\
h_2\\
h_1 \end{array}\right) = \widetilde R 
\left(\begin{array}{c}
\phi_1^0\\
\tilde{\phi}_2^0\\
\tilde{\phi}_3^0 \end{array}\right)
 = \widetilde R 
\left(\begin{array}{c}
\phi_1^0\\
c_{23}\,\phi_2^0-s_{23}\, a^0\\
s_{23}\,\phi_2^0+c_{23}\, a^0 \end{array}\right)  \ .
\ee
$\theta_{23}$ will be  important when discussing $CP$-conservation. 

Recall $\phi_1^0$  carries the full SM VEV and  exact alignment  is when $\phi_1^0$ coincides with a mass eigenstate. We choose to align $\phi_1^0$  with $h_1$, which can be achieved by setting  $c_{13}=0$ and $\theta_{13}={\pi}/2$ in Eq.~(\ref{eq:RatationMatix}). We also impose the ordering, $m_{h_1} \le m_{h_2} \le m_{h_3}$ so that $m_{h_1}=125$ GeV.

Small departures from alignment can  be parametrized by writing $\theta_{13} = {\pi}/2 + \epsilon$, $\epsilon \ll 1$, 
\begin{align}
\label{eq:approxalign}
 \widetilde R =\left(
\begin{array}{ccc}
-\epsilon\, c_{12} & -s_{12} & -c_{12}(1-\epsilon^2/2)  \\
-\epsilon\, s_{12} & c_{12} & -s_{12}(1-\epsilon^{2}/2) \\
1-\epsilon^{2}/2 & 0 & -\epsilon
\end{array}\right) \ .
\end{align}
Equation~(\ref{eq:condz2}) remains the same after we changing $\{Z_5, Z_6, Z_7\} $ into  $\{\tilde Z_5, \tilde Z_6, \tilde Z_7\}$.  We can use Eq.~(\ref{eq:condz2}) to eliminate $Z_2$ and ${\rm Im}[\tilde Z_7]$ and choose nine input parameters  $\{v,  m_{h_1}, m_{h_2}, m_{h_3}, m_{H^\pm},\theta_{12},\theta_{13}, Z_3, {\rm Re}[\tilde{Z}_7]\}$. Some important relations are, in the approximate alignment limit,
\begin{align}
{\rm Re}[\tilde{Z_5}]&=\frac{1}{v^2}\left[  c_{2\theta_{12}} \left( m^2_{h_2} -m^2_{h_3} \right)\right.\nonumber\\
& \left. + \epsilon^2 \left( m^2_{h_3} c^2_{12} + m^2_{h_2} s^2_{12} - m^2_{h_1}  \right)\right]\ , \\
{\rm Im}[\tilde{Z_5} ]&=\frac{1}{v^2}s_{2\theta_{12}}\left( 1-\frac{\epsilon^2}2 \right) \left( m^2_{h_2}- m^2_{h_3} \right)\, , \label{eq:imz5app} \\
{\rm Re}[\tilde{Z_6} ]&= \frac{\epsilon }{2v^2} s_{2\theta_{12}} \left( m^2_{h_3} - m^2_{h_2}\right)\ , \\
{\rm Im}[\tilde{Z_6} ]&= \frac{\epsilon }{v^2} \left( m^2_{h_1} -m^2_{h_2}c^2_{12} - m^2_{h_3}s^2_{12}\right) \label{eq:imz6app} \ ,\\
g_{h_1h_2h_3}&= \epsilon\, v\ {\rm Re}[\tilde{Z}_7 e^{-2i\theta_{12}}]\ . \label{eq:gh1h2h3} 
\end{align}
From the above we see that the mass splitting between $h_3$ and $h_2$ is determined at  leading order in $\epsilon$ by $\Delta m_{23}^2 \equiv (m_{h_3}^2- m_{h_2}^2)=  v^2| Z_5 |$. Therefore, in general, an ${\cal O}(v^2)$ splitting can be achieved with $|Z_5| \sim {\cal O}(1)$. Further, the CPV coupling $g_{h_1h_2h_3}$  is nonzero away from exact alignment and for nonzero ${\rm Re} [\tilde Z_7e^{-2i \theta_{12}} ]$.  Hence the decay $(h_3\to h_2 h_1)$ may be achieved for reasonable choices of parameters, which however are constrained from LHC and EDM constraints, as will be discussed later.
In the $\mathbb{Z}_2$ basis, where each field in the model has a well-defined $\mathbb{Z}_2$ charge, the Yukawa interactions must also respect the $\mathbb{Z}_2$ invariance, which necessitates assigning  $\mathbb{Z}_2$ charges to SM fermions as well \cite{Branco:1985aq,Lavoura:1994yu}. Two distinct possibilities  exist in the literature, leading to type I \cite{Haber:1978jt,Hall:1981bc} and type II \cite{Donoghue:1978cj,Hall:1981bc} models which  differ by interchanging $\tan\beta$ with $\cot\beta$ in the Yukawa couplings. Importantly $\tan\beta$ is a derived parameter~\cite{Barger:1989fj,Boto:2020wyf} which strongly depends on the mass spectrum for type II 2HDM. In the left panel of Fig.~\ref{f:tb} we show contours of $\tan\beta$ in the $m_{h_2}$ -- $m_{h_3}$ plane. For our parameter region of interest, $\tan\beta\sim1$  except when $m_{h_2}$ and $m_{h_3}$ are degenerate. 
We focus on type II models with $\tan\beta \sim {\cal O}(1)$, in which region type I and type II models have similar Yukawa couplings.

%%%%%%%%%%%%%%%%%%%%%%%%%%%%%%%%%%%%%%%

\section{Two $CP$-conserving Limits} 

The condition for $CP$ invariance in Eq.~(\ref{eq:cpconserve}) can be realized as follows \cite{Boto:2020wyf,Gunion:2005ja}:
\begin{align}
\label{eq:cpc1}
{\rm CPC1}:&\ \ {\rm Im}[\tilde{Z}_5] = {\rm Im}[\tilde{Z}_6] = {\rm Im}[\tilde{Z}_7] = 0  \ , \\
{\rm CPC2}:&\  \ {\rm Im}[\tilde{Z}_5 ] = {\rm Re}[\tilde{Z}_6 ] = {\rm Re}[\tilde{Z}_7 ] = 0 \ .
\label{eq:cpc2}
\end{align}

In CPC1, ${\widetilde {\cal M}^2}$ in Eq.~(\ref{eq:m2def}) is block-diagonal; ${\widetilde{\cal M}_{13}^2}={\widetilde{\cal M}_{23}^2}=0$, in which case $\phi_1^0$ and $\tilde{\phi}_2^0$ defined in Eq.~(\ref{eq:wideRmixing1}) are $CP$-even and can mix in general, whereas $\tilde{\phi}_3^0$ is $CP$-odd. This can be achieved by $\theta_{23}=0$ so that $\tilde{\phi}_3^0 = a^0$ in Eq.~(\ref{eq:wideRmixing1}). Further, neither of the two $CP$-even states can mix with the $CP$-odd state. From Eq.~(\ref{eq:RatationMatix}) we see $\theta_{13}$ controls the mixing between $\phi_1^0$ and $\tilde{\phi}_3^0$, which implies  $\theta_{13}=\pi/2$ in the $CP$-conserving limit.  This coincides with  the exact alignment limit $\epsilon = 0$.  The mixing between $\tilde{\phi}_2^0$ and $\tilde{\phi}_3^0$ is dictated by $\theta_{12}$ and can be removed by $\theta_{12}=0$ or $\pi/2$, which corresponds to $h_3 = a^0$ or $h_2 = a^0$, respectively. Therefore, CPC1 is reached by
 \begin{align}
 \theta_{13}=\frac{\pi}{2} \, ,   \theta_{23}= 0 \, ,  \theta_{12}= \{0, \pi/2\}\,  , {\rm Im}[{\tilde Z}_7] ={\rm Im}[{Z}_7]= 0 \  \label{eq:cpc1para}.
\end{align}

One sees from Eqs.~(\ref{eq:imz5app}) and (\ref{eq:imz6app})   that ${\rm Im}[\tilde{Z}_5] = {\rm Im}[\tilde{Z}_6]=0$ under the choice of parameters in Eq.~(\ref{eq:cpc1para}). It can be further checked that fermionic couplings of the mass eigenstates follow from their $CP$-property and the EDM constraints vanish as expected \cite{lowshahwang}.

In CPC2, ${\widetilde{\cal M}_{12}^2}={\widetilde{\cal M}_{23}^2}=0$ and ${\widetilde {\cal M}^2}$ is again block-diagonal. In this case $\phi_1^0$ can mix with $\tilde{\phi}_3^0$, since they are both $CP$-even.  The $CP$-odd state is $\tilde{\phi}_2^0$. Referring back to Eq.~(\ref{eq:wideRmixing1}) we see that this requires $\theta_{23}=\pi/2$. In contrast to the CPC1 scenario, the mixing angle $\theta_{13}$, which controls alignment, can now be arbitrary. Turning off mixing between $\tilde{\phi}_2^0$ and $\tilde{\phi}_3^0$ again implies $\theta_{12}=0$ or $\pi/2$. Hence CPC2 is represented by:\begin{equation}
\theta_{23}= {\pi}/2 \ ,  \theta_{12}= \{0, \pi/2\}\ ,  \;  {\rm Re}[{\tilde Z}_7] ={\rm Im}[{ Z}_7]  = 0 \ \label{eq:cpc2para}.
\end{equation}

Again one can check  that ${\rm Im}[\tilde{Z}_5] = {\rm Re}[\tilde{Z}_6 ] =0$ and couplings of the mass eigenstates to the fermions behave as expected from their $CP$ quantum numbers. 

There is an important distinction between these two scenarios. In CPC1 the $CP$-conserving limit coincides with the alignment limit because  misalignment introduces a small $CP$-odd component to the SM-like Higgs boson. Then the stringent EDM limits on CPV  also constrain the misalignment, $\epsilon\sim {\cal O}(10^{-4})$, thereby forcing the 125 GeV Higgs to be almost exactly SM-like \cite{lowshahwang}. This is consistent with the findings in Refs.~\cite{Grzadkowski:2014ada,Li:2015yla,Grzadkowski:2018ohf}. To the contrary, in CPC2 the SM-like Higgs boson only contains a $CP$-even non-SM-like component. Therefore EDM limits do not 
constrain misalignment.\footnote{We emphasize that this statement concerns the EDM constraints on the alignment. It was pointed out in Ref.~\cite{Grzadkowski:2014ada} that the $\mathbb{Z}_2$ condition in Eq.~(\ref{eq:condz2}) would force CPV to vanish in the exact alignment limit.}

%%%%%%%%%%%%%%%%%%%%%%%%%%%%%%%%%%%%%%%
%%%%%%%%%%%%%%%%%%%%%%%%%%%%%%%%%%%%%%%
\begin{figure}[t]	
          \includegraphics[width=0.65 \columnwidth]{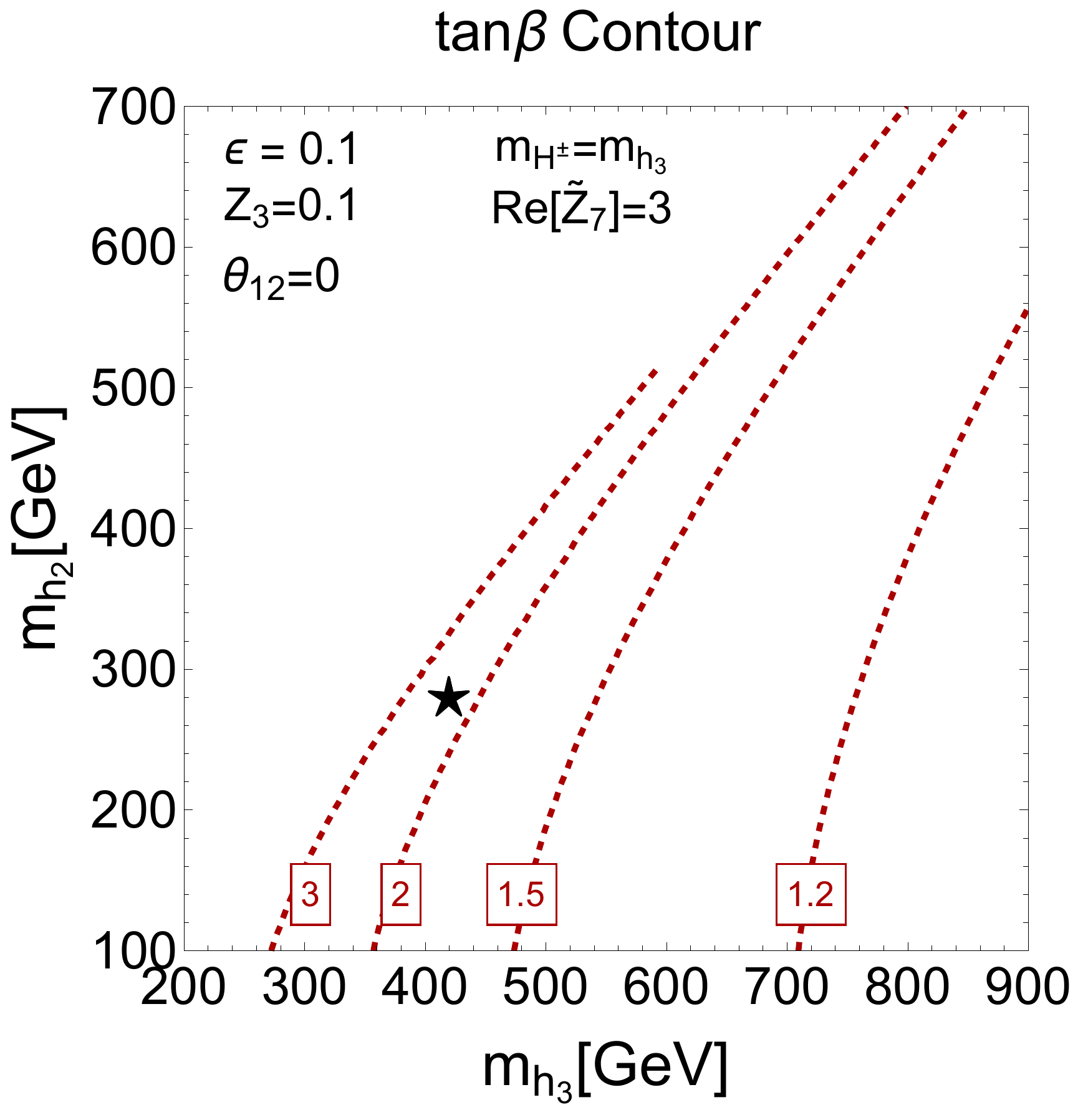}
   	\caption{  $\tan\beta$ contours in the $m_{h_2}$ - $m_{h_3}$ plane. The relevant parameters are specified in Eq.~(\ref{eq:bench}). Stars denote our benchmark point.}
	\label{f:tb1}
\end{figure}
%%%%%%%%%%%%%%%%%%%%%%%%%%%%%%%%%%%%%%%
%%%%%%%%%%%%%%%%%%%%%%%%%%%%%%%%%%%%%%%

Equations.~(\ref{eq:cpc1}) and (\ref{eq:cpc2}) also make it clear that there are two sources of CPV  in 2HDM; $\tilde{Z}_5$ and $\tilde{Z}_6$ enter into the scalar mass-squared matrix in Eq.~(\ref{eq:m2def}), while $\tilde{Z}_7$ does not. When ${\rm Im}[\tilde{Z}_5]={\rm Im}[\tilde{Z}_6]=0$ or ${\rm Im}[\tilde{Z}_5]={\rm Re}[\tilde{Z}_6]=0$, there is no CPV  in the scalar mixing matrix and each mass eigenstate $h_i$ is also a $CP$-eigenstate: two are $CP$-even and one is $CP$-odd. In this case, CPV could still be present through nonzero ${\rm Re}[\tilde{Z}_7]$ or ${\rm Im}[\tilde{Z}_7]$ and will manifest  in the bosonic interactions of the Higgs bosons. 
In light of these considerations,  we construct
a benchmark which interpolates between the CPC1 and CPC2 limits,
\begin{align}
&\!\!\! \!\!\!\{ Z_3, {\rm Re }[\tilde{Z}_7], \theta_{12}, \theta_{23}, \epsilon\}=\{ 0.1, 3,{\pi}/{2}, 1.23, 0.1 \} \nonumber , \\
&\!\!\! \!\!\!\{m_{h_3}, m_{h_2}, m_{H^\pm}\}= \{420 , 280 , 420 \} {\rm\ GeV}\ .
\label{eq:bench}
\end{align}

In Fig.~\ref{f:tb1} we show the $\tan\beta$ contours on the $m_{h_3}-m_{h_2}$ plane, for the region of parameter space close to our benchmark; our benchmark has $\tan\beta \sim 2.3$. With these parameters, ${h_1}$ is mostly $CP$-even, while ${h_2}$ and ${h_3}$ are $CP$-mixed states.  In our benchmark the charged Higgs and $h_3$ are degenerate in mass so as to be consistent with precision electroweak measurements, which include the oblique parameters $S$, $T$ and $U$ \cite{,Hessenberger:2016atw,Haber:2010bw,Funk:2011ad}. Conventional wisdom has it that  a charged Higgs lighter than 800 GeV   is constrained by $b\to s\gamma$ measurements \cite{Misiak:2017bgg,Misiak:2020vlo}. However, more recent results \cite{Bernlochner:2020jlt} argued that the theoretical uncertainty leaves more room for the new physics contribution. So in our analysis, we set the charged Higgs mass at 420 GeV, which is considered safe for the $b\to s \gamma$ measurement \cite{Altmannshofer:2020shb}.

%%%%%%%%%%%%%%%%%%%%%%%%%%%%%%%%%%%%%%%

\section{LHC/EDM Constraints} 

%%%%%%%%%%%%%%%%%%%%%%%%%%%%%%%%%%%%%%%
%%%%%%%%%%%%%%%%%%%%%%%%%%%%%%%%%%%%%%%
\begin{figure}[t]	
        \includegraphics[width=0.65 \columnwidth]{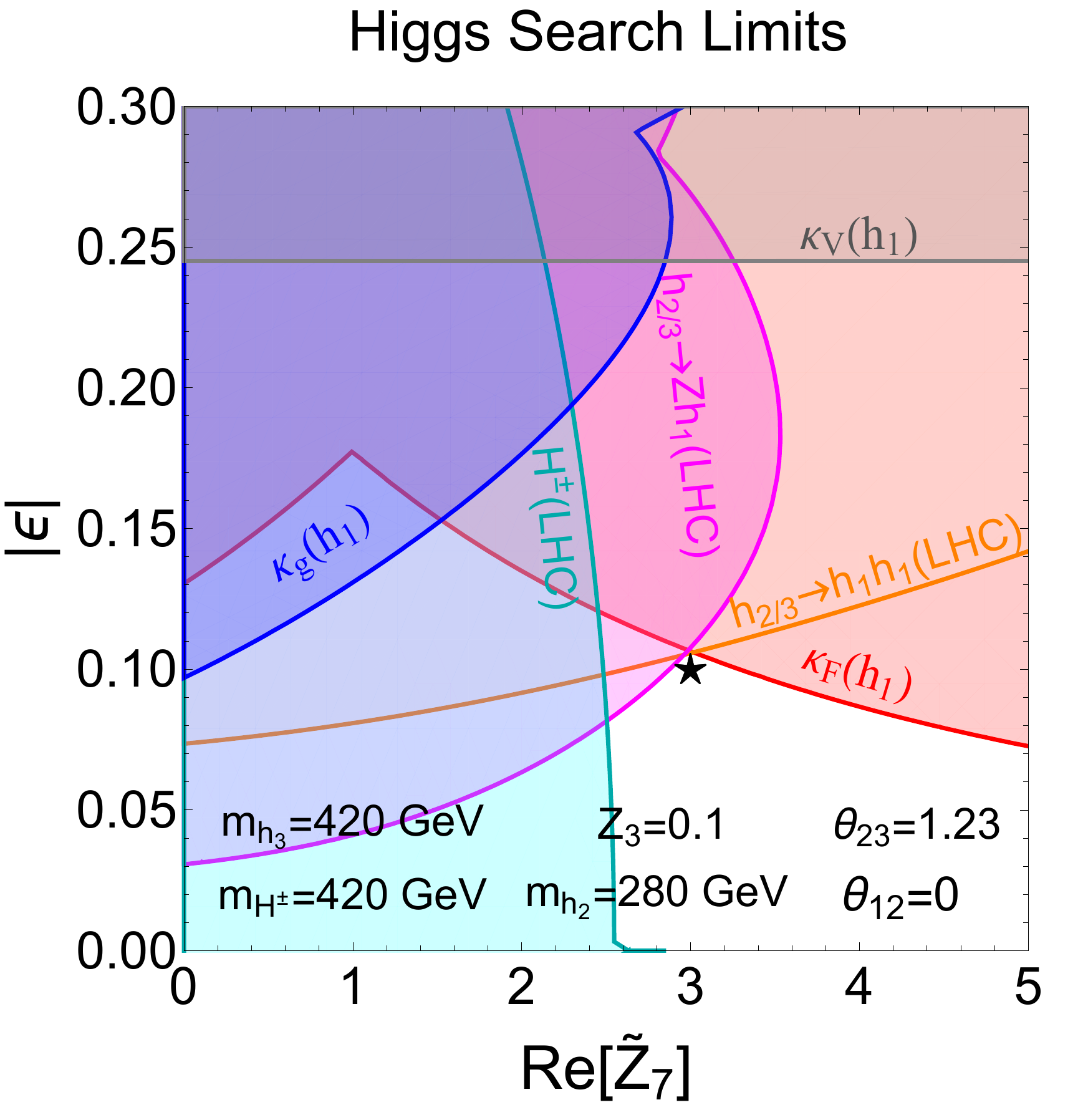}
   	\caption{ LHC constraints on $|\epsilon|$ from Higgs couplings with gluons ($\kappa_g$), vector bosons ($\kappa_V$), fermions ($\kappa_F$) and photons ($\kappa_\gamma$), as well as searches for $H^+ \to tb$ (cyan) , $h_{2/3} \to Z h_1$(magenta) and $h_{2/3} \to h_1h_1$ (orange). Stars denote our benchmark point.}
	\label{f:tb}
\end{figure}
%%%%%%%%%%%%%%%%%%%%%%%%%%%%%%%%%%%%%%%
%%%%%%%%%%%%%%%%%%%%%%%%%%%%%%%%%%%%%%%

In  Fig.~\ref{f:tb} we show the LHC constraints on $|\epsilon|$ and ${\rm Re}[\tilde{Z}_7]$.  
  For Higgs coupling measurements we use  recent results from both ATLAS \cite{ATL-PHYS-PUB-2018-054,PhysRevD.101.012002} and CMS~\cite{CMS-PAS-HIG-17-031}, which constrain $\kappa_i=g_i^{\rm measured}/g_i^{\rm SM}, i=g, V, F, \gamma$. Blue, gray, red and green shaded regions correspond to regions excluded by constraints coming from $\kappa_g$, $\kappa_V$, $\kappa_F$, and $\kappa_\gamma$, respectively. The cyan shaded region is excluded due to searches for $H^+ \to tb$~\cite{Sirunyan:2019arl,ATLAS-CONF-2020-039},  which requires $\tan \beta \geq 2$ and is satisfied by our benchmark point, $\tan \beta=2.3$. 
For $m_{h_3(h_2)}=420(280)$ GeV the experimental limit from double Higgs production ~\cite{Sirunyan:2018two} is shown as the orange shaded region in Fig.~\ref{f:tb} and the limit from $h_3/h_2\to Zh_1$ search \cite{ATLAS:2017xel,CMS-PAS-HIG-18-005} is given by the magenta shaded region in Fig.~\ref{f:tb}. We also checked that LHC limits on heavy Higgs decays to $t\bar{t}$ final states  \cite{Sirunyan:2019wph} are not relevant for our benchmark.

%%%%%%%%%%%%%%%%%%%%%%%%%%%%%%%%%%%%%%%
%%%%%%%%%%%%%%%%%%%%%%%%%%%%%%%%%%%%%%%
\begin{figure}[h!]	
             \includegraphics[width=0.8\columnwidth]{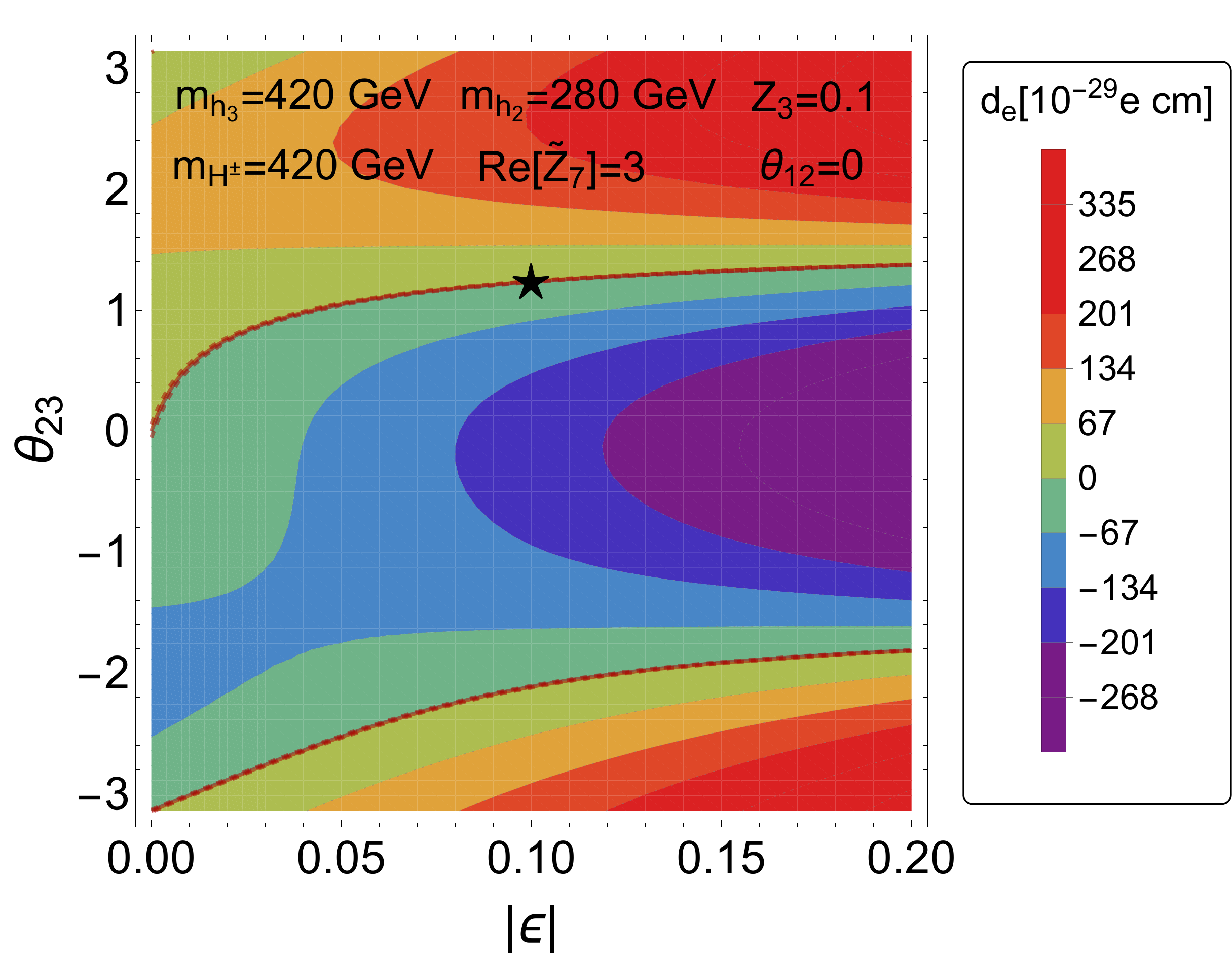}
           \includegraphics[width=0.8 \columnwidth]{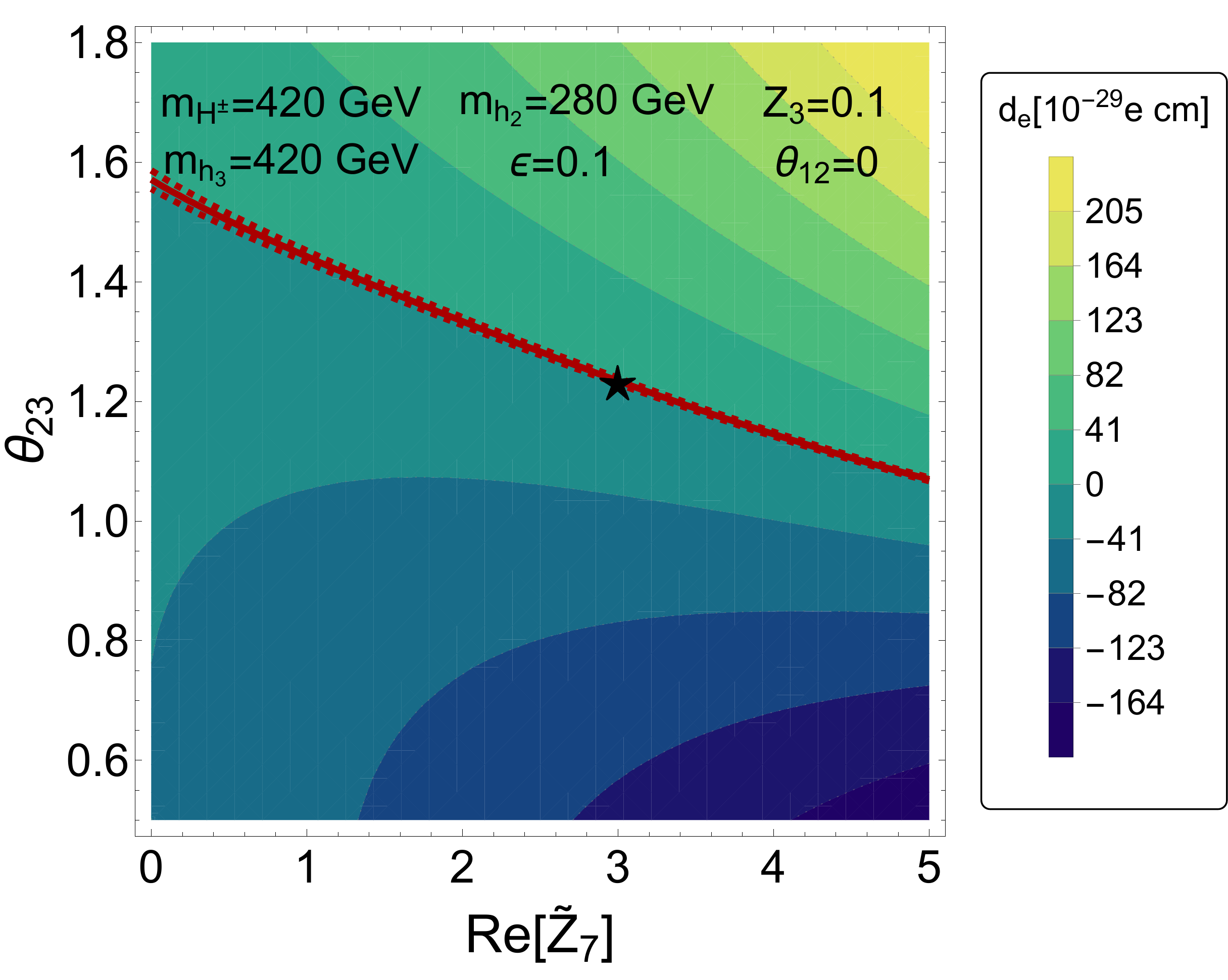} 
        	\caption{ \footnotesize{Contours for  eEDM~($d_e$) in $\theta_{23}$ vs. $|\epsilon|$~(top), and $\mathrm{Re} [\tilde{Z}_7] $~(bottom) plane. Only regions within the dashed red lines are experimentally allowed $|d_e|<1.1\times 10^{-29}{\rm e~ cm}~(90\%{\rm CL})$~\cite{Andreev:2018ayy}. Thick red line denotes $|d_e|=0$. 
	Note different scales for the left/right axes and legends. Stars denote our benchmark point.}}
	\label{f:edm}
\end{figure}
%%%%%%%%%%%%%%%%%%%%%%%%%%%%%%%%%%%%%%%
%%%%%%%%%%%%%%%%%%%%%%%%%%%%%%%%%%%%%%%

%%%%%%%%%%%%%%%%%%%%%%%%%%%%%%%%%%%%%%%

In our analysis, we consider both constraints from the electron EDM (eEDM)  \cite{Cheung:2014oaa,Jung:2013hka,Andreev:2018ayy}  and neutron EDM (nEDM)  \cite{Abel:2020gbr}. The most recent constraints are 
\begin{align}
|d_n|&<1.8\times 10^{-26} {\rm e~cm}~(90\% {\rm C.L.}) \\ \nonumber
|d_e|&<1.1\times 10^{-29} {\rm e~cm}~(90\% {\rm C.L.}). 
\end{align}
The dominant contribution for both eEDM and nEDM are the two-loop Barr-Zee(BZ) diagrams \cite{Barr:1990vd,Pilaftsis:1999qt,Abe:2013qla,Heinemeyer:2013tqa,Inoue:2014nva,Bian:2014zka,Egana-Ugrinovic:2018fpy}. The  BZ diagrams to $d_f(f=e,~d,~{\rm and}~u)$ and $d_q^C (q=d~{\rm and}~u)$ contain contributions from fermion-loops, Higgs boson-loops, and gauge boson-loops \cite{Kanemura:2020ibp}
\begin{align}
d_f=d_f({\rm fermion})+d_f( {\rm Higgs})+d_f({\rm gauge}),
\end{align}
and each contribution includes 
\begin{align}
d_f(X)=d_f^\gamma(X)+d_f^Z(X)+d_f^W(X).
\end{align}
For the nEDM, the relevant formula for $d_n$ \cite{Abe:2013qla} is
\begin{align}
d_n=0.79d_d-0.2d_u+\frac{e}{g_s}\left( 0.59 d_d^C+0.3d_u^C\right),
\end{align}
where $g_s$ is the QCD gauge coupling constant. Both fermion-loops and gauge boson-loops contributions are related to the $CP$ property of the Yukawa interactions, which are parametrized by $\theta_{23}, \theta_{12}$ and $\epsilon$. The Higgs-loops contributions are both related to the $CP$ property of Yukawa interaction and the coupling of $g_{H^\pm H^\mp h_j}$, which not only depends on $\epsilon$, $\theta_{23}$, $\theta_{12}$, but also depends on  ${\rm Re}[\tilde Z_7]$. After consider both eEDM and nEDM, we found that the eEDM constraints are stronger than those from the neutron EDM, so we only show the relevant plots for eEDM. In Fig.~\ref{f:edm}   contours for the eEDM and the experimental constraints on the most relevant parameters are shown: $\theta_{23}$ vs $\epsilon$ (left) and ${\rm Re}[\tilde Z_7]$ (right). The solid red line denotes $d_e=0$, while the dashed red lines bound the experimentally allowed region $|d_e|<1.1\times 10^{-29}{\rm e~ cm}~(90\%{\rm C.L.})$~\cite{Andreev:2018ayy}. We fix the mass spectrum as for the LHC constraints, and again choose $\theta_{12}=0$. While not shown, EDM constraints are minimized when the masses are degenerate \cite{Boto:2020wyf}. However, regardless of the mass spectrum, eEDM constraints severely limit the CPV components of the mass eigenstates. This can be seen from the limits on $d_e$ tracking the behavior expected from our analysis of CPC1 and CPC2. Small values of $\theta_{23}$~(CPC1 limit) can only be obtained for small values of $|\epsilon|$, but for $|\theta_{23}|\sim \pi/2$~(CPC2), $\epsilon$ is effectively unconstrained. Further, small values of ${\rm Re}[\tilde Z_7]$ are obtained for values of $\theta_{23}\sim \pi/2$~(CPC2 limit), but larger values are allowed as $\theta_{23}$ decreases. Additionally, we see that in regions far from CPC1 and CPC2, $d_e$ can be 0 due to cancellations between various contributions. This is the region where our benchmark resides.

%%%%%%%%%%%%%%%%%%%%%%%%%%%%%%%%%%%%%%%

\section{Collider Phenomenology} 

With the generically small CPV components allowed in the mass eigenstates due to experimental constraints, directly probing the $CP$ nature of the mass eigenstates will be challenging. However, the decay  $(h_3\to h_2 h_1)$ could provide a smoking gun signature for CPV in 2HDMs. If kinematically accessible, this signal is maximized for maximum possible misalignment $\epsilon$ and largest possible ${\rm Re}[\tilde Z_7]$~[cf. Eq.~(\ref{eq:gh1h2h3})], as allowed from LHC and where eEDM constraints are minimized. Further, we are interested in  the possibility of both additional Higgs bosons being within reach of the LHC, which motivates the benchmark presented
in Eq.~(\ref{eq:bench}).

%%%%%%%%%%%%%%%%%%%%%%%%%%%%%%%%%%%%%%%
%%%%%%%%%%%%%%%%%%%%%%%%%%%%%%%%%%%%%%%
\begin{figure}[t]	
         \includegraphics[width=0.65 \columnwidth]{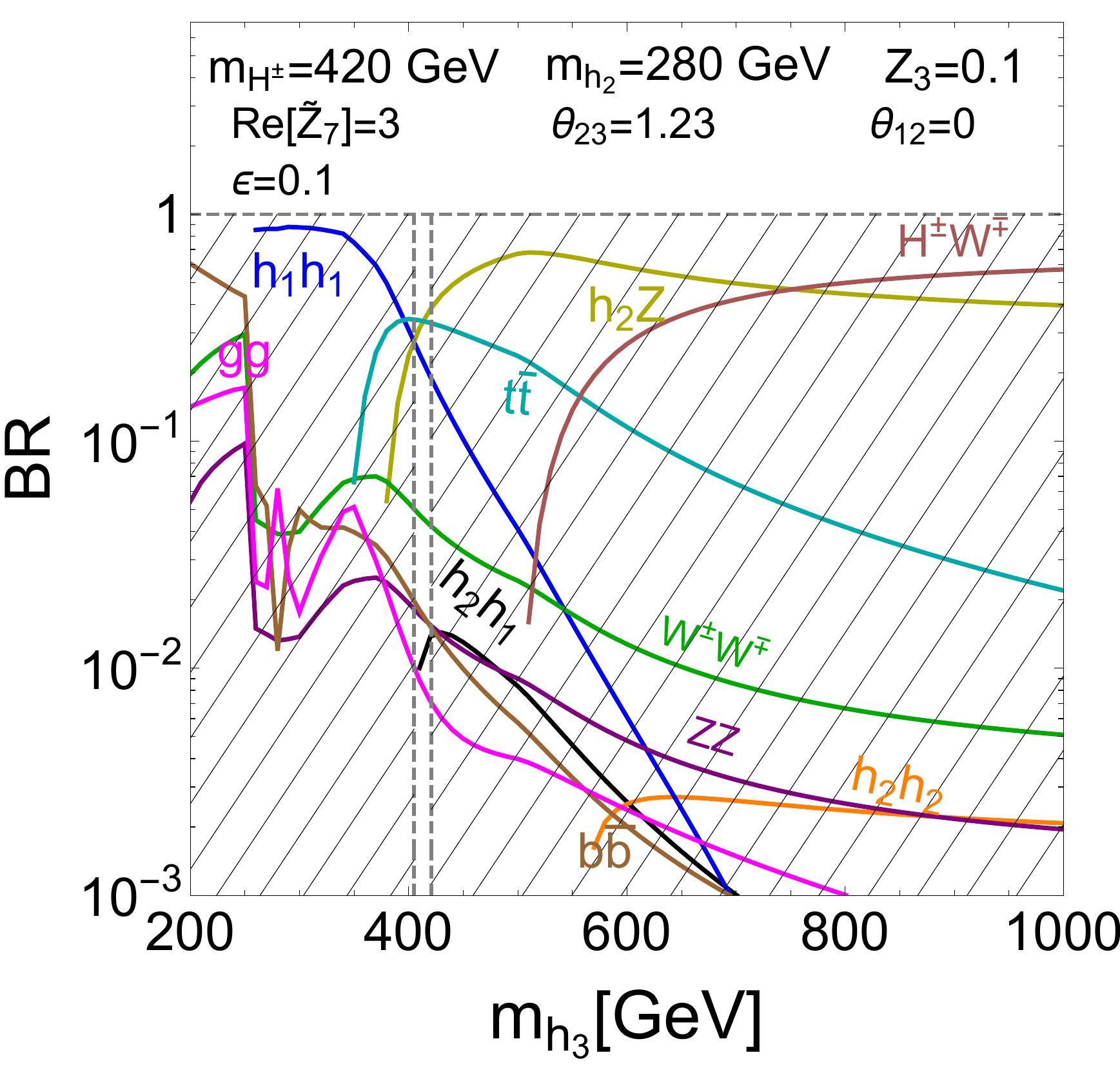} 
          \includegraphics[width=0.65\columnwidth]{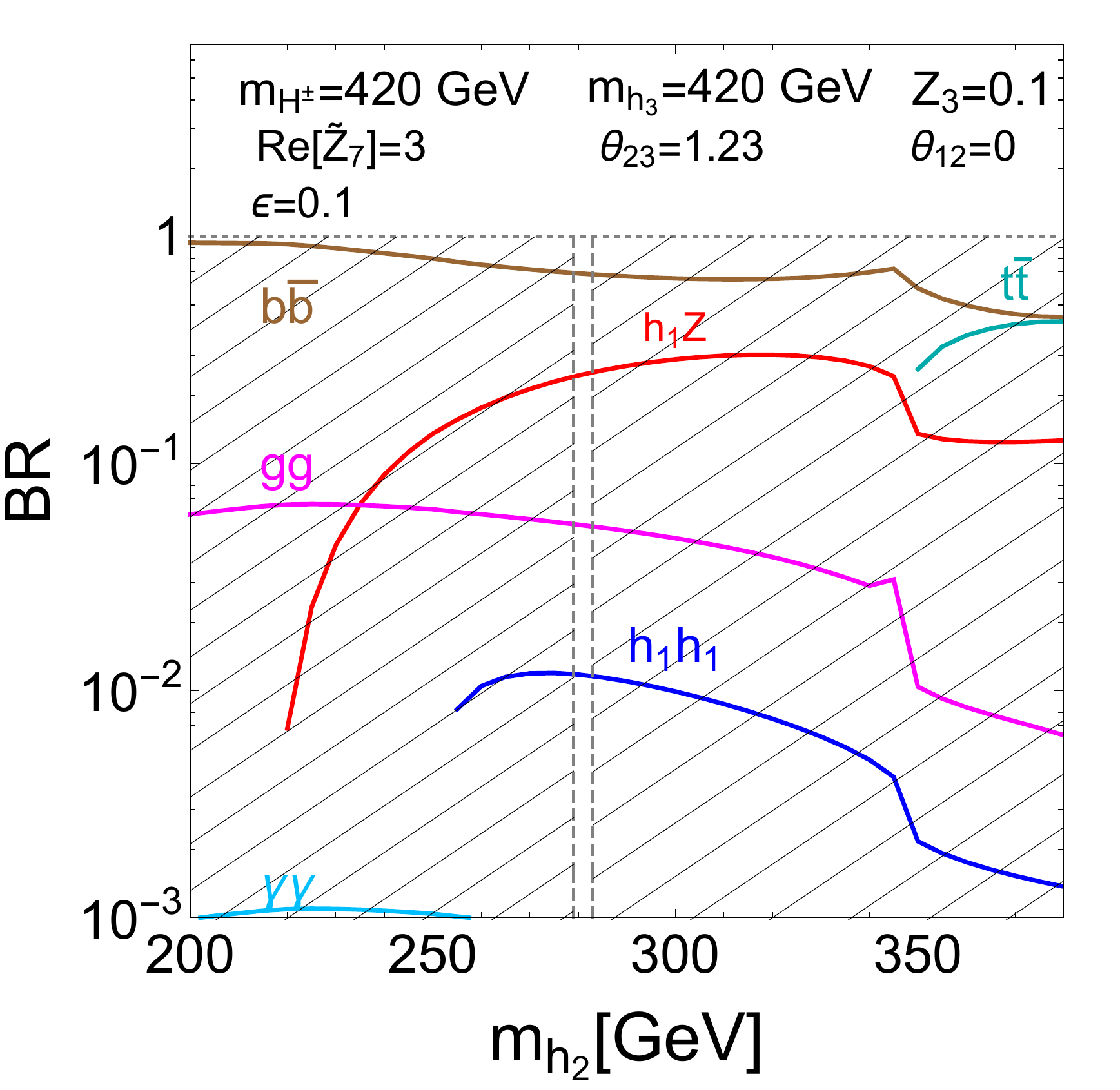} 
        	\caption{Branching ratios for $h_3$~(top) and $h_2$~(bottom) for the listed parameters. Gray dashed lines denote mass spectra in tension with eEDM constraints for chosen set of parameters. }
	\label{BR-H1}
\end{figure}
%%%%%%%%%%%%%%%%%%%%%%%%%%%%%%%%%%%%%%%
%%%%%%%%%%%%%%%%%%%%%%%%%%%%%%%%%%%%%%%

Fig.~\ref{BR-H1} shows the  branching ratios of $h_3$~(top panel) and $h_2$~(bottom panel). 
Gray hatching denotes mass spectra in tension with eEDM constraints.One particular decay mode we would like to focus on is the CPV Higgs-to-Higgs decay, $h_3\to h_2h_1$, which in our benchmark has a branching fraction ${\rm BR}(h_3~\to~h_2h_1)\sim 1.5\%$.  As can be seen from Eq.~(\ref{eq:gh1h2h3}), the coupling controlling this decay is $CP$-violating. Under the assumption that there are two $CP$-even and one $CP$-odd scalars in 2HDM, it is easy to see that such a decay mode is forbidden if $CP$ is conserved. 

From Fig.~\ref{BR-H1} we  also see that $h_2$ can decay into $h_1 h_1$ and $h_1 Z$, giving rise to $h_3\to h_1h_1h_1$ and $h_3\to h_1h_1Z$ final states via $h_3\to h_2h_1$. The main production channel for both $h_2$ and $h_3$ is gluon fusion. At the $\sqrt{s}=13$ TeV LHC \cite{Heinemeyer:2013tqa}: 
\begin{align}
\sigma(gg \to h_2) \simeq 5.8~{\rm pb}\;,  \qquad 
\sigma(gg \to h_3) \simeq 2.7~{\rm pb}\; .
\end{align}
The large production rate for $h_3$ stems from its sizable $CP$-odd component.
Therefore, for an integrated luminosity of $\mathcal{L}=3000~\rm{fb}^{-1}$,
we will  have approximately $500$ CPV triple Higgs events from
 $h_3\to h_1 h_1 h_1$, which is a smoking gun signature of $CP$-violation in C2HDMs. 
 
The decay modes of $h_3$, include $h_3\to h_2 Z$ and $h_3\to h_1 h_2$, will gives rise to the final state $h_1ZZ$ and $h_1 h_1 Z$, from $h_2\to h_1Z$, $h_2\to Z Z$ and $h_2\to h_1h_1$.  The event rates for $h_1h_1 Z$ and $h_1ZZ$ at the HL-LHC with $\mathcal{L}$=3000 fb$^{-1}$ are $2\times 10^4$ and $2\times 10^5$.  

Alternatively, $h_2$ could be produced directly through the gluon fusion mechanism. In this case, simultaneous observations of $h_2\to h_1h_1$ and $h_2\to h_1Z$ would be an unambiguous signal of $CP$-violation. The events rates for $h_1 h_1$ and $h_1Z$ are $4 \times 10^4$ and $9\times 10^5$ at the HL-LHC.

These triboson signatures have not been searched for at the LHC and represent excellent opportunities to pursue CPV in 2HDMs at a high-energy collider. Moreover, the relatively light mass of $h_2$ and its decays into two 125 GeV Higgs bosons also imply a significant discovery potential in the near future.

\section{Conclusion}  

Motivated by the SM-like nature of the 125 GeV Higgs and null searches for new particles at the LHC, we  present a systematic study of Higgs alignment and CPV in  C2HDMs and distinguish two distinct sources of CPV in the scalar sector. The outcome is the construction of a new $CP$ violating scenario where additional Higgs bosons could be light, below 500 GeV, and stringent EDM limits and current collider searches may still be evaded.

In particular, we propose a smoking gun signal of CPV in C2HDMs in the $h_1h_2h_3$ coupling through the Higgs-to-Higgs decays, $(h_3\to h_2h_1\to 3 h_1)$, without resorting to the challenging measurements of kinematic distributions. The existence of this decay in C2HDMs is indicative of CPV and the final state in three 125 GeV Higgs bosons is quite distinct, which has not been searched for at the LHC. A ballpark estimate demonstrates the great potential for discovery at the high-luminosity LHC.

\section*{Acknowledgment} 

We would like to thank Marcela Carena, Kai-Feng Chen, Cheng-Wei Chiang, Howie Haber, George Wei-Shu Hou, Shinya Kanemura, Jia Liu, and Carlos Wagner for useful discussions and comments. N.R.S. is supported by U.S. Department of Energy under Contracts No. DE-SC0021497. I.L. is supported in part by the U.S. Department of Energy under contracts No. DE-AC02-06CH11357 at Argonne and No. DE-SC0010143 at Northwestern. X.P.W. is supported by NSFC under Grant No.12005009. I.L. also acknowledges the hospitality from the National Center for Theoretical Sciences at National Tsing Hua University and National Taiwan University in Taiwan where part of this work was performed.

\bibliography{CPX2HDM_ref}

\end{document}